\documentclass[aps,showpacs,prd,twocolumn]{revtex4}
\usepackage{amssymb,amsfonts,amsmath,bm}

\begin{document}

\title{Chiral symmetry, the angular content of the vector current in QED and
QCD, and the holographic description of hadrons}

\author{L. Ya. Glozman}
\affiliation{Institute for Physics, Theoretical Physics Branch, University of Graz, Universit\"atsplatz 5, A-8010 Graz, Austria}

\author{A. V. Nefediev}
\affiliation{Institute of Theoretical and Experimental Physics, 117218, B.Cheremushkinskaya 25, Moscow, Russia}

\begin{abstract}
We perform a general chiral symmetry and unitarity based
analysis of a local process of the fermion-antifermion
creation from the vacuum by a high-energy photon as well as an explicit 
partial wave analysis of the vector current in QED and QCD.
It turns out that such a local process proceeds necessarily via a 
certain superposition of the $S$- and $D$-wave contributions.
These constraints from chiral symmetry and unitarity are confronted
then with the well-known theoretical and experimental results on
$e^+e^-\to\gamma\to e^+e^-$, $e^+e^-\to\gamma\to \mu^+\mu^-$, 
and $e^+e^-\to\gamma\to q\bar{q}$
in the ultrarelativistic limit. It is shown that these well-known
results are consistent with the $S+D$-wave structure of the vertex
and are inconsistent with the pure $S$-wave interpretation of the vertex.
Then a free quark loop in the $1^{--}$ channel, representing the leading
term in the Operator Product Expansion, contains both $S$-wave and $D$-wave
contributions. This fact 
rules out the possibility that there is only one radial trajectory
for the $\rho$-mesons with the fixed $S$-wave content. It also implies that all 
holographic models
that assume a pure $S$-wave content of the $\rho$-meson have to fail to satisfy
the matching conditions at the ultraviolet border $z=0$.
\end{abstract}

\maketitle
\section {Introduction}

Symmetries are known to be a powerful method to analyse and understand 
physics. They suggest stringent constraints on dynamics in many physical 
systems. In particle physics chiral symmetry is one of the most 
important ingredients. If chiral symmetry is broken spontaneously then 
there are Nambu--Goldstone bosons, and many results of the soft physics 
can be understood as being due to the interaction of the Goldstone 
bosons with other particles, without any necessity of knowing the 
microscopical details of these interactions. A prominent example of this 
kind is the chiral perturbation theory. If, on the contrary, chiral 
symmetry is unbroken or is restored for a particular reason, then this 
manifest chiral symmetry implies stringent constraints for the 
dynamics, structure, 
and spectra of physical states. In particular, we demonstrate
in this paper that such  
constraints exist to the dynamics of various high-energy
processes in QED and QCD, 
including such elementary ones as the electron--positron, 
muon--antimuon, or quark--antiquark pairs creation by high-energy
photons. These 
vertices are believed to be generally local (point-like) and to 
have an $S$-wave structure due to their locality (see, for example,
Ref.~\cite{Swave1}). Here we demonstrate 
that the point-like vertex of creation of a free fermion--antifermion 
pair from the vacuum by an energetic photon has a more complicated structure and 
that chiral symmetry 
together with unitarity require a fixed superposition of the $S$- and 
$D$-waves in such a local vertex. This implies that the leading term of the
Operator Product Expansion (OPE) in the vector-isovector channel contains
both $S$- and $D$-wave contributions which rules out a possibility that
there is only one radial $\rho$-meson trajectory populated by $S$-wave
$\rho$'s.
It also implies that holographic models assuming a pure $S$-wave content of
the $\rho$-mesons violate the standard AdS/CFT matching conditions at the
ultraviolet border $z=0$.

\begin{table}[t]
\caption{The complete set of $q\bar{q}$ states classified according to 
the chiral basis. The sign $\leftrightarrow$ indicates the states 
belonging to the same representation.}
\begin{ruledtabular}
\begin{tabular}{cccc}
$R$&$J=0$&$J=1,3,\ldots$&$J=2,4,\ldots$\\
\hline
$(0,0)$&\bf ---&$0J^{++} \leftrightarrow 0J^{--}$&$0J^{--}  \leftrightarrow 0J^{++}$\\
$(1/2,1/2)_a$&$1J^{-+}\leftrightarrow 0J^{++}$&$1J^{+-} \leftrightarrow 0J^{--}$&$1J^{-+} \leftrightarrow 0J^{++}$\\
$(1/2,1/2)_b$&$1J^{++}\leftrightarrow 0J^{-+}$&$1J^{--} \leftrightarrow 0J^{+-}$&$1J^{++} \leftrightarrow 0J^{-+}$\\
$(0,1) \oplus (1,0)$&\bf ---&$1J^{--} \leftrightarrow 1J^{++}$&$1J^{++} \leftrightarrow 1J^{--}$\\
\end{tabular}\label{t1}
\end{ruledtabular}
\end{table}

\section{Chiral symmetry and the angular momentum content of the vector current}

We start with quite general chiral symmetry based arguments
for a fermion--antifermion system.
If chiral $SU(2)_L \times SU(2)_R$ symmetry is unbroken, systems with
the fermion--antifermion valence degree of freedom, for example
quark--antiquark states, fall into chiral 
multiplets --- see Table~\ref{t1} for a complete classification of
such multiplets \cite{G2,G3}. 
Here the index $R$ determines a representation of the $SU(2)_L \times SU(2)_R$
($R=$ $(0,0)$, $(1/2,1/2)_a$, $(1/2,1/2)_b$, or $(0,1)+(1,0)$). Then all states
are uniquely specified by the set of quantum numbers $\{R;IJ^{PC}\}$
where we use the standard notations for the isospin $I$, total spin $J$ as well
as for
the spatial and charge parities $PC$. The chiral basis 
$\{R;IJ^{PC}\}$ is obviously consistent with the Poincar{\'e} invariance.
The sign $\leftrightarrow$ indicates that both given states are
members of a particular chiral multiplet, that is they transform into each other
upon $SU(2)_L \times SU(2)_R$ and therefore the corresponding physical states
must be degenerate.

The chiral basis can be related, through a unitary transformation, to the 
$\{I;{}^{2S+1}L_J\}$ basis  in the centre-of-momentum frame \cite{GN123}:
\begin{eqnarray}
|R;IJ^{PC}\rangle&=&\sum_{L}\sum_{\lambda_q\lambda_{\bar q}}
\chi_{\lambda_q \lambda_{\bar q}}^{RPI}\nonumber\\[-2mm]
\label{5}\\[-2mm]
&\times&\sqrt{\frac{2L+1}{2J+1}}
C_{\frac12\lambda _q\frac12-\lambda_{\bar q}}^{S\Lambda}
C_{L0S\Lambda}^{J\Lambda}|I;{}^{2S+1}L_J\rangle,\nonumber
\end{eqnarray}
where the summation is implied in helicities of the fermion $\lambda_q$
and antifermion $\lambda_{\bar q}$  as well as 
in the orbital angular momenta  $L$'s such 
that $(-1)^{L+1}=P$. Coefficients $\chi_{\lambda_q \lambda_{\bar q}}^{RPI}$
can be extracted from Table~2 of Ref.~\cite{GN123}.
It follows immediately from Eq.~(\ref{5}) that 
every state in the chiral basis is a fixed (prescribed by chiral 
symmetry and unitarity) superposition of allowed states in the
$\{I;{}^{2S+1}L_J\}$ 
basis. Basis states $\{R;IJ^{PC}\}$
are normalised  and the variables are  two
angles that specify a direction of the relative momentum of two
quarks in the center-of-momentum frame.
For instance, there are two kinds of the vector particles with
the quantum numbers of the $\rho$-meson, which are represented by two orthogonal
fixed combinations of the $S$- and $D$-waves:
\begin{eqnarray*}
\displaystyle |(0,1)+(1,0);1 ~ 1^{--}\rangle&=&\sqrt{\frac23}|1;
{}^3S_1\rangle+\sqrt{\frac13}|1;{}^3D_1\rangle,\\
\displaystyle |(1/2,1/2)_b;1 ~ 1^{--}\rangle&=&\sqrt{\frac13}|1;
{}^3S_1\rangle-\sqrt{\frac23}|1;{}^3D_1\rangle.
\end{eqnarray*}
The same decomposition obviously applies to the spatial components
of the  composite operators  such as 
$\bar{q}\gamma^i{\bm\tau}q$ and $\bar{q}\sigma^{0i}{\bm\tau}q$,
because they create from the vacuum  the $1,1^{--}$ states and transform as 
$(0,1)+(1,0)$, in the former case, and as $(1/2,1/2)_b$, in the latter
\cite{G2,G3,CJ}.

These are constraints from the $SU_L(2)\times SU_R(2)$ symmetry. A natural
question
arises, what are constraints from the $U_L(1)\times U_R(1)= 
U_A(1)\times U_V(1)$? It turns out that this symmetry group
adds nothing new and all the coefficients for the $S$- and $D$-waves are
exactly the same. One can obtain this easily from the explicit unitary
transformation
similar to that from Eq.~(\ref{5}). It can also be seen in a simpler way,
however. Indeed, the state $|(0,1)+(1,0); 1~ 1^{--}\rangle$
and its $SU_L(2)\times SU_R(2)$ partner $|(0,1)+(1,0); 1~ 1^{++}\rangle$
are both invariant with respect to the $U(1)_A$ (the same is true, of course,
for the vector and the axial vector currents). Similarly
states (or interpolators) from the the $(0,0)$ representation of
the $SU_L(2)\times SU_R(2)$ group also transform into themselves (that is
they are invariant) upon $U(1)_A$. On the other hand, states (interpolators)
from the $(1/2,1/2)_a$ representation transform upon $U(1)_A$ into
states (interpolators) with the same isospin but with the opposite
spatial parity, which belong to the $(1/2,1/2)_b$ representation, and vise
versa. All these properties require that the $U(1)_A$ symmetry together with
the unitarity prescribe exactly the same $S+D$-wave decompositions
of the vector and pseudotensor currents. Hence, in the one-flavour case, 
properly normalised vector and pseudotensor interpolators in the
centre-of-momentum frame are also given by
\begin{eqnarray}
\displaystyle |\bar{q}\gamma^i q \rangle&=&\sqrt{\frac23}|
{}^3S_1\rangle+\sqrt{\frac13}|{}^3D_1\rangle\label{veccur},\\
\displaystyle |\bar{q}\sigma^{0i} q \rangle&=&\sqrt{\frac13}|
{}^3S_1\rangle-\sqrt{\frac23}|{}^3D_1\rangle.
\end{eqnarray}
Here $|\bar{q}\gamma^i q \rangle$ and $|\bar{q}\sigma^{0i} q \rangle$
are symbolic notations for the states
$|\frac {1}{\sqrt 2}(\bar R R + \bar L L); ~ 1^{--}\rangle$ and
$|\frac {1}{\sqrt 2}(\bar R L + \bar L R); ~ 1^{--}\rangle$,
respectively. In such a way the $U(1)_L \times U(1)_R$ representation
is uniquely specified.

\section {$e^+e^- \to \mu^+\mu^-$ as a test of the
angular momentum content of the vector current}

The above analysis is not specific for 
QCD and is therefore applicable to any fermion--antifermion pair
in the chiral or ultrarelativistic limit. So this
partial wave decomposition, in particular that of the vector current,  
holds also for QED if one can neglect masses of fermions, such as electrons,
muons, and so on, that is in the ultrarelativistic limit.

A key point is that this decomposition is uniquely specified by
the chiral ($U(1)_V \times U(1)_A$ or $SU(2)_L \times SU(2)_R$)
symmetry and unitarity. A process of creation from the vacuum of a
fermion--antifermion pair by a high-energy photon  
in QED and in QCD is believed to be local. At very short
distances chiral symmetry is unbroken in QCD due to the
asymptotic freedom, so these constraints from chiral symmetry should apply both
to QED processes like $e^+e^-\to\gamma\to e^+e^-$, 
$e^+e^-\to\gamma\to \mu^+\mu^-$ in the ultrarelativistic limit
as well as to the process of the electron--positron annihilation into two
jets in QCD, $e^+e^-\to\gamma\to q\bar{q}$. In the leading order
the axial anomaly is irrelevant so the prediction from the chiral
$U(1)_A$ symmetry is valid. One can
anticipate that due to locality these processes proceed  through the
$S$-wave only.
This expectation is in conflict with the
partial wave decomposition prescribed by chiral symmetry and unitarity,
however. Then a key question is whether the partial wave decomposition 
above meets (or contradicts) the well-known theoretical and experimental
results for these reactions. In the following we demonstrate that indeed
the local process of creation from the vacuum of a fermion--antifermion pair
by a high-energy photon proceeds via a superposition
of the $S$- and $D$-waves.

Let us consider the process of a $e^+e^-$ pair annihilation to a pair of 
fermions $f\bar{f}$ with the electric charge $Q$. In the lowest order of 
the perturbation theory this process is simply $e^+e^-\to\gamma\to 
f\bar{f}$, so that the angular dependence of the differential 
cross-section as well as the value of the total cross-section will allow 
us to make definite conclusions concerning the structure of the vector 
current responsible for this process. We stick to the ultrarelativistic 
limit to preserve chiral symmetry. We stress that predictions of chiral
symmetry cannot be correct for the low-energy processes. One has:
\begin{equation}
\frac{d\sigma}{do}=\frac{\alpha^2 Q^2}{16 E^2}\sum_{\{\rm
polarisations\}}|{\cal M}|^2.
\label{crsec}
\end{equation}
Here $E$ is the initial energy of the electron (or positron) in the
centre-of-momentum frame, while the sum goes over polarisations 
of all four fermions involved.  It is easy to verify then 
that
\begin{equation}
\frac{d\sigma}{do}=\frac{\alpha^2 Q^2}{16 E^2}(1+\cos^2\theta),
\quad\sigma=\frac{\pi\alpha^2 Q^2}{3E^2},
\label{sigtot}
\end{equation}
which is the well-known result (for $Q=1$) and is a subject of textbooks
(see, for example, \cite{ahiber}). 
In this expression the angle $\theta$ is defined by the three-momenta 
of the produced particles with respect to the direction of motion of colliding
particles in the centre-of-mass system.

Our task now is to clarify whether this result is consistent with the
chiral symmetry prediction (\ref{veccur}) or it can be obtained assuming a pure
$S$-wave structure of the vertex. The latter possibility is ruled out however
by the $1+\cos^2\theta$ angular dependence of the squared amplitude. Indeed,
$\cos^2\theta$ clearly indicates a contribution of the $D$-wave.

 To check the
former possibility we have to look into the structure of the amplitude
$\sum|{\cal M}|^2$. 
Traditionally this amplitude squared is obtained using the standard
algebra of $\gamma$-matrices, that amounts to taking trace of the product of a
number of matrices. In this way the partial wave content of the amplitude is
not explicit and is obscured. However, exactly the same result,
\begin{equation}
\sum_{\{\rm polarisations\}}|{\cal M}|^2 =1+\cos^2\theta,
\label{ampl}
\end{equation}
can be obtained with the technique of vector spherical harmonics
${\bm Y}_{JM}^L(\theta,\varphi)$ and  ${\bm
Y}_{JM}^{(\lambda)}(\theta,\varphi)$ (see, for example, Ref.~\cite{V}).
The harmonics ${\bm Y}_{JM}^L(\theta,\varphi)$ are eigenfunctions of the
operators ${\bm J}^2$, $J_z$, ${\bm L}^2$, and ${\bm S}^2$, 
where ${\bm J}={\bm L}+{\bm S}$ and ${\bm S}$ is the spin operator for $S=1$.
The harmonics ${\bm Y}_{JM}^{(\lambda)}(\theta,\varphi)$ are not eigenfunctions
of the operator ${\bm L}^2$ but, instead, they are transverse (for
$\lambda=1,0$) or longitudinal (for $\lambda=-1$) with respect to
${\bm n}(\theta,\varphi)$.
 
There exists an identity:
\begin{equation}
1+\cos^2\theta =\frac{16\pi}{3}|{\bm Y}_{11}^{(1)}(\theta,\varphi)|^2.
\label{eq1}
\end{equation}
In turn,
\begin{equation}
{\bm Y}_{11}^{(1)}(\theta,\varphi)=\sqrt{\frac23}
{\bm Y}_{11}^0(\theta,\varphi)+\sqrt{\frac13}{\bm Y}_{11}^2(\theta,\varphi),
\label{eq2}
\end{equation}
which has exactly the structure of Eq.~(\ref{veccur}).
Averaging over polarisations is equivalent to averaging over the
spin projection $M$. We note that the amplitude  squared with $M=-1$
is the same as the amplitude with $M=1$ given in Eq.~(\ref{eq1}).
The spin projection $M=0$ is
not allowed in a system of two massless fermions with 
the quantum numbers of $\bar{q}\gamma^i q$.

Consequently we recover that the amplitude squared is
precisely the superposition of the $S$- and $D$-waves with
the coefficients required by chiral symmetry and unitarity.
It is obvious then that the $D$-wave contribution cannot
be omitted. Indeed, if we remove the $D$-wave contribution, then
the total cross-section $\sigma$ will become only $2/3$ of the
correct result (\ref{sigtot}). This is because the $S$- and $D$-waves do
not interfere in the total cross-section.

We have demonstrated therefore an exact equivalence
of the chiral symmetry plus unitarity based result 
and the standard results for the reactions $e^+e^-\to\gamma\to 
f\bar{f}$ in the ultrarelativistic limit. This gives in fact a theoretical 
and experimental
verification of the chiral symmetry plus unitarity constraint
for the high-energy $\gamma\rightarrow f\bar{f}$ vertex.
We stress  that the above analysis does not apply to 
situations when chiral symmetry is broken. In the latter case chiral symmetry 
breaking implies a smaller weight of the $D$-wave in the amplitude.

\section{Implications of the $S+D$-wave structure of the local 
$\gamma \to f\bar{f}$ vertex}

To our best knowledge this $S+D$-wave interpretation of the local 
$\gamma \to f\bar{f}$ vertex in the chiral or ultrarelativistic
limit is new and deserves some attention. Certainly it helps to give a
proper physical interpretation to many basic processes in QED and QCD
(and in general, in particle physics), not only those considered here. 
Below we consider some of the most straightforward implications.

\subsection{$L$ is not a conserved quantum number in mesons}

Chiral symmetry is broken in the infrared, so that the
quark-antiquark component of the $\rho$-meson is a superposition
of the $(0,1)+(1,0)$ and $(1/2,1/2)_b$ representations with
the ratio close to $\sqrt{2}$. This implies that, in the infrared,
the wave function of the $\rho$-meson is almost a pure $S$-wave, as it follows 
from the model-independent and manifestly gauge-invariant
dynamical lattice calculations in QCD \cite{L}. However, at
high resolution scales $Q^2\to\infty$, QCD
is conformal and the pseudotensor interpolator decouples
from the $\rho$-meson (it has a nonzero anomalous dimension). Then,
in the deep ultraviolet the angular momentum content of the
$\rho$-meson is determined solely by the $(0,1)+(1,0)$ representation
with the fixed superposition of the $S$- and $D$-waves. 
This implies that models which assume a fixed (conserved)
angular momentum $L=0$ for the $\rho$-meson are not realistic and appear to be
in odds with QCD.

\subsection {The OPE and the $\rho$-meson radial trajectory}

The total cross-section (\ref{sigtot}) is given by the imaginary part of the
fermionic vacuum polarisation by a photon. It can be continued analytically
to the deep Euclidean region $Q^2=-q^2$ and produces a well-known
leading term in the OPE for the two-point correlator of the vector
current,
\begin{equation}
\Pi(Q^2)=-\frac{N_c}{12\pi^2}\ln\left(\frac{Q^2}{\mu^2}\right).
\label{OPE}
\end{equation}
The OPE has been used in many works to constrain the spectrum of mesons ---
see Ref.~\cite{S} for a review. In the large-$N_c$
limit all excited mesons are stable and, in the time-like domain
$s=q^2$, along the cut,
\begin{equation}
\Pi(s)=-\sum_{n=1}^\infty\frac{f^2_n}{s-m_n^2} 
\label{OPE2}
\end{equation}

A typical assumption made was that the radially excited $\rho$-mesons lie on
the Regge trajectory, $m_n^2 \propto n$, and, in the semiclassical regime,
they have
a fixed $L=0$ \cite{SV}. Then, under a certain assumption on the
constants $f_n$, the result (9) can be reproduced from the sum in Eq. (10).

However, as was demonstrated above, the leading term in the OPE is a superposition of
the $S$- and $D$-waves with the weights 2/3 and 1/3, respectively.
Therefore the spectrum of excited $\rho$-mesons cannot consist of only one
linear Regge trajectory populated by $S$-wave mesons. If excited $\rho$-mesons
are assumed to be pure $S$-waves, the OPE cannot match the spectrum.  

\subsection {Holographic description of mesons}

Recently AdS/QCD holographic models of hadrons have become very popular.
The heart of the holographic description of hadrons, according to
the AdS/CFT dictionary, is matching of the operators composed
of free quarks and gluons with the required quantum numbers 
with their duals in AdS at
the ultraviolet boundary $z=0$, that corresponds to the ultraviolet
resolution scale $Q^2\to\infty$. There is no chiral symmetry
breaking at $z=0$ and consequently composite QCD operators must satisfy the
angular momentum decomposition prescribed by Eq.~(\ref{5}). In particular, the
vector current, which is typically used in the holographic description of the
$\rho$-mesons, is a certain superposition of the $S$- and $D$-waves.
Consequently, its dual in the bulk of the 5D AdS space which depends on the
coordinate $z$ and which satisfies a certain differential equation \cite{ADS},
must satisfy exactly the same $S+D$-waves decomposition at the ultraviolet
boundary $z\to 0$. We are not aware of any demonstration that it does. Even
more, there are some popular holographic models \cite{bdt} which use the vector
current for the $\rho$-mesons and assume at the same time a fixed orbital
angular momentum $L=0$ for them, even at $z \rightarrow 0$.
 Certainly these models are in
conflict with chiral symmetry and prescriptions of the dictionary.
\vspace*{-1mm}

\acknowledgments
The authors thank S. Brodsky and G. de Teramond for discussions.
L. Ya. G. acknowledges support of the Austrian Science Fund through grant 
P19168-N16. Work of A. N. was supported by the Federal Agency for Atomic 
Energy of Russian Federation by 
the grants RFFI-09-02-00629a, RFFI-09-02-91342-NNIOa, DFG-436 RUS 
113 /991/0-1(R), NSh-843.2006.2, PTDC/FIS/70843/2006-Fi\-si\-ca, and by 
the non-profit ``Dynasty'' foundation and ICFPM.

\end{document}